\documentclass[sigconf,nonacm]{acmart}
\usepackage{mdframed}

\newenvironment{summary}
{\begin{mdframed}\noindent }
{\end{mdframed}}

\AtBeginDocument{%
  }

\acmConference[KDD '26]{Agentic Software Engineering (SE 3.0)}{August 9, 2026}{Jeju, South Korea}

\graphicspath{{./images/}}

\begin{document}

\title{Early Adoption of Agentic Coding Tools by GitHub Projects
}

\author{Maliha Noushin Raida}
\affiliation{%
  \institution{Rochester Institute of Technology}
  \city{Rochester, NY}
  \country{USA}
}

\author{Daqing Hou}
\affiliation{%
  \institution{Rochester Institute of Technology}
  \city{Rochester, NY}
  \country{USA }}

\renewcommand{\shortauthors}{Raida and Hou}

\begin{abstract}

Agentic coding tools are increasingly capable of generating and submitting pull requests (PRs) to software projects, introducing new forms of human-agent collaboration in software development. While prior studies have examined PR-level outcomes of agent-generated contributions, less is known about how agentic coding tools are adopted and managed at the project level. In this paper, we analyze 25,264 agentic PRs from 2,361 popular GitHub repositories to investigate (1) the adoption of agentic coding tools, (2) project-level agentic PR productivity, and (3) human-agent collaboration patterns.
Our results show that the median repository generates only one to two agentic PRs during a three-month period, indicating that intensive adoption remains concentrated in a relatively small subset of projects. At the same time, small projects (1-5 contributors) exhibit substantially higher participation ratios and average levels of agentic PR activity than medium-sized and large projects. We also observe substantial variation in project-level agentic PR productivity. While a small number of projects exceed an industry-reported estimate of 36 PRs per participant during the three-month observation period, most projects remain below this threshold. Finally, human-agent collaboration is dominated by a single-human oversight model, in which one developer both reviews and/or modifies the agent's contributions, while multi-human collaboration patterns remain comparatively uncommon.
These findings provide early empirical evidence on how open-source projects organize human oversight around agentic coding tools and suggest that the successful integration of agent-generated contributions depends not only on advances in agent capabilities but also on the human and organizational processes that govern their use.
Because this study captures an early snapshot of agent adoption, future work should continue to track how adoption patterns evolve over time.

\end{abstract}




\maketitle
\begingroup
\renewcommand\thefootnote{}
\footnotetext{Accepted for presentation at the \textit{KDD 2026 Workshop on Agentic Software Engineering (SE 3.0)}, August 9, 2026, Jeju, South Korea.}
\addtocounter{footnote}{-1}
\endgroup

\section{Introduction}
Agentic coding tools are rapidly transitioning from completion-style assistants to semi-autonomous teammates that can independently generate, test, and submit pull requests (PRs) to large-scale software projects~\cite{hassan2024towards}. As these tools become integrated into development workflows, they raise foundational questions about how human developers coordinate with and oversee agents, and to what extent agents are being adopted by projects of varying team sizes.

Recent empirical work using the AIDev~\cite{li2025aiteammates} and AIDev-pop datasets~\cite{li2025-AIDev-pop-huggaingface} has begun to document the growing presence of agentic PRs across thousands of GitHub repositories. Complementary studies have examined the productivity implications of AI-assisted software development and the adoption of AI coding agents in open-source ecosystems~\cite{10.1145/3772318.3790850,khan2026adoption}, providing early evidence that AI-generated contributions are becoming an increasingly important part of modern software development workflows. Prior studies have also examined how humans and agents divide labor across PR lifecycles~\cite{jo2026collaborator,watanabe2025use} and how PR characteristics influence merge outcomes~\cite{ehsani2026ai, yoshioka2026let,peralta2026agentic}. However, most existing studies focus on PR-level outcomes, such as quality, acceptance, and labor allocation, rather than project-level adoption patterns.  

Understanding project-level adoption is important because the effectiveness of agentic coding tools depends not only on the capabilities of the agents themselves but also on the human oversight and organizational processes that support their use. Projects with different team sizes and contributor structures may adopt and manage agentic contributions in different ways. 

In this paper, we address this gap by analyzing the adoption of agentic coding tools across open-source projects using  AIDev-pop focusing on three widely used AI coding agents: \textit{Copilot}, \textit{Codex}, and \textit{Claude Code}. We restrict our analysis to merged and closed PRs created during a three-month period (May, June, and July, 2025). 

This paper addresses the following research questions:

\begin{itemize}
    \item \textbf{RQ1:} To what extent have agentic coding tools been adopted across GitHub projects?
    \item \textbf{RQ2:} How does agentic PR productivity vary across GitHub projects?
    \item \textbf{RQ3:} How do human-agent collaboration patterns in agentic PRs differ across projects of varying team sizes?
\end{itemize}

To answer these questions, we analyze project-level adoption, agentic PR productivity, and human participation patterns across 2,361 repositories and 25,264 agentic PRs.

Our results show that, although agentic coding tools have been adopted across thousands of projects, intensive adoption remains concentrated in a limited number of projects. The median repository generates only one to two agentic PRs during the three-month observation period. At the same time, small projects exhibit substantially higher participation ratios and average levels of agentic PR activity than medium-sized and large projects and involve a larger proportion of their contributors in agentic workflows. We also find substantial variation in project-level agentic PR productivity, with only a small number of projects exceeding the industry-reported activity level of 36 PRs per participant during the three-month observation period. Finally, human-agent collaboration is dominated by a single-human oversight model, in which one developer both reviews and modifies the agent's contributions, while more distributed review practices remain comparatively uncommon.

Overall, our findings provide early empirical evidence on how open-source projects adopt  agentic coding tools and organize human oversight around agentic PRs. They also highlight the importance of human review capacity and project governance in supporting the sustainable adoption of agentic software engineering practices.

\section{Related Work}
Recent research has characterized AI coding agents as a new stage in software engineering, moving beyond code-completion assistants toward autonomous teammates capable of generating, testing, and submitting pull requests (PRs) ~\cite{hassan2024towards}. The AIDev dataset~\cite{li2025aiteammates} and its AIDev-pop~\cite{li2025-AIDev-pop-huggaingface} subset of popular repositories provide large-scale evidence that agent-generated PRs are increasingly common in open-source software development and offer a foundation for studying human-agent collaboration in real-world development workflows.

Several studies have investigated the adoption of AI coding agents and characteristics of agentic PRs. Prior work has investigated the adoption of AI coding agents in open-source projects~\cite{khan2026adoption}, factors associated with agentic PR acceptance and rejection~\cite{ehsani2026ai, peralta2026agentic}, differences between human-authored and agent-generated PRs~\cite{yoshioka2026let}, and empirical characteristics of agentic pull requests on GitHub~\cite{watanabe2025use}. Other studies have explored the impact of AI coding tools on software development productivity and development practices~\cite{asdaque2026novice}. Collectively, these studies demonstrate the growing use of AI coding agents while highlighting the variability of their outcomes across projects and development contexts.

Another line of research focuses on human-agent collaboration in software engineering. Recent studies have examined how humans and coding agents divide work throughout the PR lifecycle, ranging from assistant-like interactions to more collaborative forms of participation~\cite{jo2026collaborator}. This study provide important insights into the roles played by humans and agents during code generation, review, and integration activities. Complementing this work, Chen et al.~\cite{10.1145/3772318.3790850} present the first controlled study comparing developer interactions with copilot-style assistants and more autonomous coding agents, finding that agents can surpass copilot's assistance by completing tasks that humans may not have accomplished alone and by reducing the effort required to finish tasks, yet challenges remain around users' ability to adequately understand agent behaviors and outputs. 

Understanding these collaborative dynamics requires considering the broader literature on software engineering productivity. The SPACE framework~\cite{10.1145/3453928} established that developer productivity is multidimensional, encompassing satisfaction, performance, activity, communication, and efficiency. Empirical studies of open-source projects further show that productivity and delivery efficiency vary substantially across projects and are strongly influenced by factors such as team size and contributor structure~\cite{SCHREIBER2026112765,10.1145/3510003.3510619}. These findings suggest that organizational context plays an important role in shaping development outcomes and may similarly influence how projects benefit from agentic coding tools.

Recent large-scale experiments have begun to quantify the productivity effects of AI-assisted development. Several studies report that AI tools can improve productivity by increasing task completion rates, reducing completion time, and enhancing output quality across a variety of professional settings~\cite{cui2024effects, doi:10.1126/science.adh2586, doi:10.1287/orsc.2025.21838}. Complementing these findings, Ziegler et
al.~\cite{10.1145/3633453} conducted a large-scale case study of GitHub
Copilot users, demonstrating that perceived productivity gains are reflected in objective usage measurements, and acceptance rate is the strongest predictor of self-reported productivity across all SPACE dimensions, with junior developers showing the largest gains. However, other studies have found more limited or inconsistent benefits in terms of task completion time from using LLMs, in software development contexts~\cite{vaithilingam2022chi}. Taken together, these findings suggest that the productivity impacts of AI tools can be substantial but remain highly dependent on task characteristics, user experience, and organizational context.

Despite these advances, prior work has largely examined agentic PRs at the level of individual contributions. What remains underexplored is how adoption, productivity, and human participation around agentic PRs differ across projects and team structures. Our work complements prior studies by examining these project-level questions through a large-scale analysis of agentic PRs from GitHub repositories. By focusing on project-level adoption, agentic PR activity, and human-agent collaboration across projects of different sizes, our study extends prior PR-level analyses with a project-level perspective on how open-source communities integrate agentic coding tools into their workflows.

\section{Methodology}
This section outlines the dataset, the identification of human participants in agentic PR workflows, the grouping of projects by team size, and the classification of human-agent participation patterns.
\subsection{Dataset}

This study uses the AIDev-pop dataset~\cite{li2025-AIDev-pop-huggaingface}, a large-scale collection of agentic PRs from popular GitHub repositories. We selected this dataset because it provides the only publicly available collections of real-world AI-generated software contributions across diverse open-source projects. Unlike benchmark datasets that evaluate coding agents on controlled programming tasks, AIDev-pop captures actual development activities involving AI coding agents and human collaborators. Its scale, repository diversity, and rich interaction records make it well suited for studying human-agent collaboration patterns, participation structures, and the adoption of AI coding agents in open-source software development.

The dataset contains 33,596 agentic PRs associated with 2,807 repositories that have at least 100 GitHub stars and were active during the study period. This study focuses on PRs generated using three widely adopted coding agents: GitHub Copilot, OpenAI Codex, and Claude Code.

AIDev-pop integrates information from multiple GitHub artifacts, including pull requests, commits, reviews, review comments, and timeline events. These artifacts enable the reconstruction of human-agent interactions throughout the PR lifecycle and support the analysis of project-level adoption and collaboration patterns.

For this study, we focus on PRs created during a three-month observation period (May-July 2025). To ensure that PR outcomes were observable, we include only PRs that were either merged or closed without merging by the time the dataset was collected. We exclude repositories for which contributor information could not be retrieved through the GitHub API, as contributor counts are required for project-size classification and participation analysis.

\begin{table}[!h]
\centering
\caption{Filtered Dataset Overview}
\label{tab:dataset_overview}
\begin{tabular}{l l}
\hline
\textbf{Attribute} & \textbf{Value} \\
\hline
Dataset & AIDev-pop \cite{li2025-AIDev-pop-huggaingface} \\
Period & May-July 2025 \\
Agentic PRs & 25,264 \\
Repositories & 2,361 \\
Commits & 291,866 \\
Agents & Copilot, Codex, Claude Code \\
Artifacts & PRs, commits, reviews, comments \\
\hline
\end{tabular}
\vspace{-3mm}
\end{table}
The filtered dataset summarized in Table~\ref{tab:dataset_overview} forms the basis of all subsequent analyses.

\subsection{Human Participants in Agentic Workflows}
To identify human participation in agentic PR workflows, we integrate multiple artifacts from the AIDev-pop dataset at the pull-request level. We first  filtered PR records from the \textit{pull\_request} table, which contains attributes such as its ID, URL, agent label, lifecycle state, and timestamps.
These filtered records include merged and closed PRs during the three months, which are used as the foundation of our analysis. These PRs are then joined with review, comment, commit,  and timeline tables to extract reviewers, committers, and actors.

To identify humans who directly modified code associated with a PR, we analyze records from the \textit{pr\_commit\_details} table. Each PR is linked to one or more commits, from which we extract information about commit authors, committers, and commit messages. A commit is classified as agent-authored if any of these fields contains identifiers associated with coding agents or automated accounts (e.g., \texttt{codex}, \texttt{claude}, \texttt{copilot}, \texttt{bot}, or \texttt{github}). After excluding agent-authored commits, the remaining distinct committers are treated as human code contributors to the PR.

We also capture review and discussion activity using records from \textit{pr\_reviews} and \textit{pr\_review\_comments}. These records identify users who formally reviewed PRs or participated in discussions. For each PR, these records are aggregated to construct a set of distinct reviewers and commenters. 

In addition, we use \textit{pr\_timeline} events to capture actors involved in PR lifecycle events, particularly \texttt{merged} and \texttt{closed} actions, which help identify users involved in resolving PRs. These events provide additional evidence of human participation in the review and integration process.

Finally, we combine committers, reviewers, commenters, and timeline actors into a unified participant set for each PR. We remove bot and agent accounts using keywords such as \texttt{[bot]}, \texttt{web-flow}, \texttt{claude}, \texttt{codex}, \texttt{copilot}, \texttt{mergify}, \texttt{sre-ci-robot}, and \texttt{cursor}. The resulting participant set represents the distinct human contributors involved in reviewing, discussing, modifying, or resolving each agentic PR. We use the size of this participant set when normalizing agentic PR activity across projects and computing project-level agentic PR productivity.

\subsection{Project Teams}
AIDev-pop does not include repository-level contributor counts. However, contributor team size is central to our analysis because it allows us to examine whether project-level  adoption and human-agent collaboration patterns vary systematically across small, medium, and large development teams.

We collected contributor information for all 2,361 repositories using the GitHub API (endpoint: /repos/owner/repo/contributors). We define contributor count as the number of distinct non-bot accounts with at least one commit to the repository's default branch, which serves as a proxy for project team size. Repositories were then stratified into three categories: small (1-5 contributors), medium (6-15 contributors), and large (16+ contributors). This resulted in 343 small projects, 456 medium projects, and 1,562 large projects.

Because all agentic PRs in our dataset were created within the same three-month observation period (May-July 2025), we use the number of human participants associated with agentic PR workflows as a normalization factor when comparing projects. Specifically, for RQ2, we operationalize project-level agentic PR productivity as the number of agentic PRs per human participant during the three-month observation period. Although agentic PRs may vary in size, complexity, and review effort, this measure provides a proxy for the volume of agentic contributions generated and managed per participant.
\begin{figure}[!t]
    \centering
    \includegraphics[width=\linewidth]{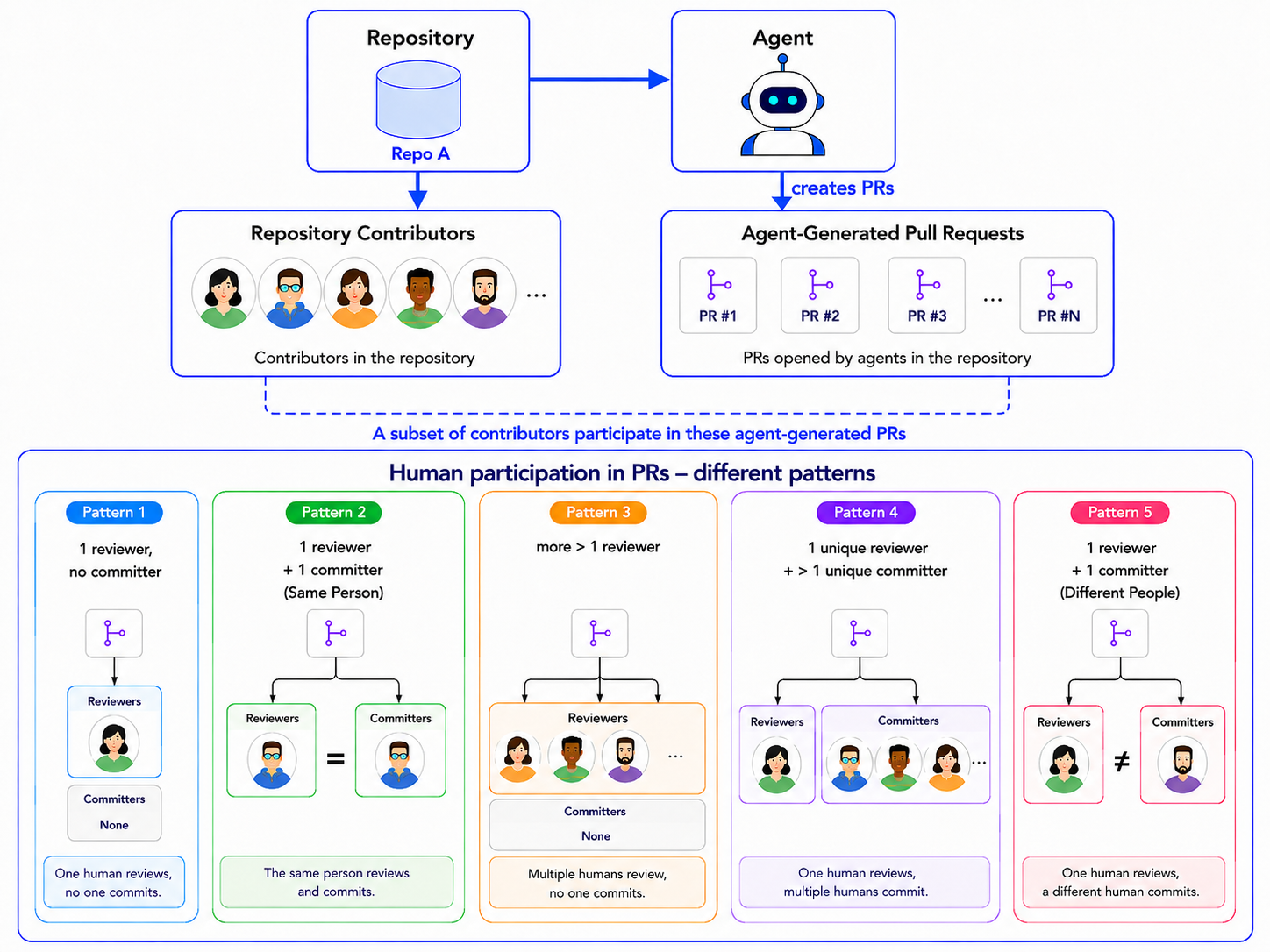}
    \caption{Different human participation patterns in agentic PRs}
\label{fig:scenarios}
\vspace*{-1.7\baselineskip}
\end{figure}

\subsection{Human Participation Patterns}
We classify every agentic PR into one of five mutually exclusive participation patterns based on the number and identity of reviewers and committers involved in each PR (Figure~\ref{fig:scenarios}):
\begin{enumerate}
    \item \textbf{1 Reviewer, No Committer:} The PR receives one human review but no human commits. The agent is the sole code contributor.

    \item \textbf{1 Reviewer + 1 Committer, Same Person:}
    One human both reviews and commits code, representing a solo-developer agentic workflow.

    \item \textbf{1 Reviewer + 1 Committer, Different People:} One human reviews and a different human commits, reflecting a formal separation of review and development responsibilities.
    \item \textbf{1 Reviewer + $>$1 Committer:} One human reviewer oversees code committed by multiple humans alongside the agent.
    \item \textbf{$>$1 Reviewer, No Committer:} Multiple humans review the PR, but no human commits code. The agent's output is reviewed but accepted as-is.

\end{enumerate}

The five patterns represent the dominant configurations observed in the dataset. Two additional collaboration patterns (i.e, \textbf{$>1$ Reviewer + 1 Committer} and \textbf{$>1$ Reviewer + $>1$ Committer}) occurred rarely in the dataset, hence were excluded from the analysis.
\section{Results}
This section reports results on project-level adoption, agentic PR productivity, and human-agent collaboration patterns across different project team sizes.

\subsection{RQ1: Adoption of Agentic Coding Tools Across Small/Medium/Large Project Teams}
To measure project-level adoption, we use two complementary indicators: (1) the human participation ratio, defined as the proportion of repository contributors who participate in at least one agentic PR, and (2) the number of agentic PRs generated per repository.

\begin{figure}[t]
\centering\includegraphics[width=\linewidth]{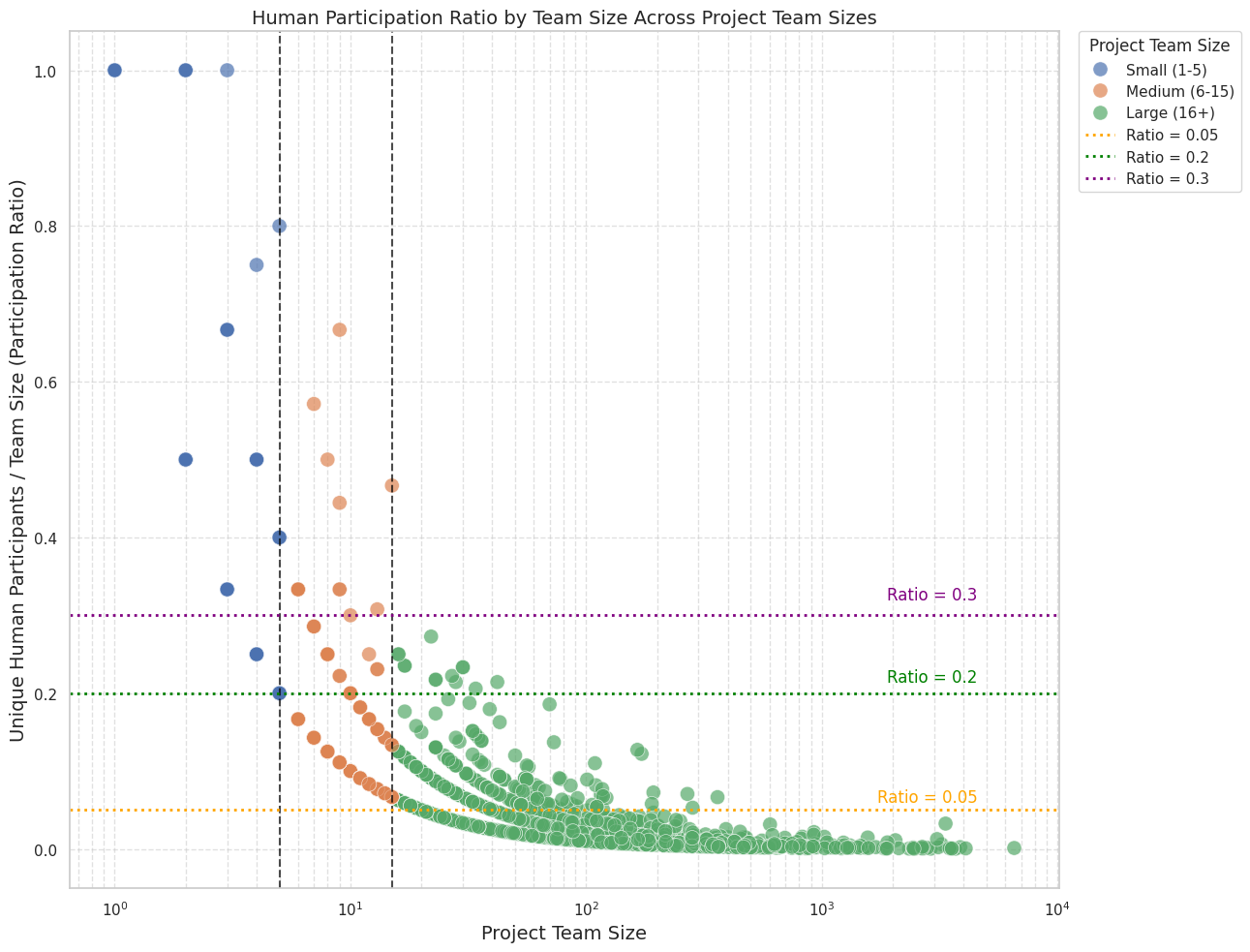}
    \caption{Project team sizes (small, medium, and large) versus human participation ratio in agentic PRs}
    \label{fig:participation_ratio}
    \vspace*{-\baselineskip}
\end{figure}

Figure~\ref{fig:participation_ratio} illustrates how broadly contributors participate in agentic workflows across projects of different sizes. Small projects exhibit the highest participation ratios, indicating that a larger share of their contributors is involved in agentic PR activities. In contrast, medium-sized and large projects have substantially lower participation ratios, suggesting that agentic workflows are concentrated among a smaller subset of contributors.

The distribution also shows that low participation ratios are common across the dataset. Among the 2,361 projects, 998 projects (42.27\%) have a participation ratio below 0.05, meaning that fewer than 5\% of their contributors participate in agentic PR workflows. Similarly, 1,657 projects (70.18\%) have participation ratios below 0.20, and 1,776 projects (75.22\%) have participation ratios below 0.30. These results indicate that, although agentic coding tools have been adopted across many projects, participation in agentic workflows is typically limited to a relatively small portion of a project's contributor base.

To evaluate whether these differences are statistically significant, we performed the Shapiro-Wilk~\cite{10.1093/biomet/52.3-4.591} test which indicated deviations from normality. Therefore, we conducted a non-parametric Kruskal-Wallis H-test~\cite{ac6c544c-0197-38bd-8c06-ec4f655ff4fd} and the results revealed a statistically significant difference in participation ratios across the three project-size categories (\(H = 1211.79\), \(p < 0.001\)). We computed eta-squared for the Kruskal-Wallis test as \(\eta_H^2 = (H - k + 1)/(n - k)\), where \(H\) is the H-statistic, \(k\) is the number of groups, and \(n\) is the total sample size. The corresponding effect size was large (\(\eta_H^2 = 0.6157\)), indicating that approximately 61.6\% of the variation in participation ratios is associated with project-size categories. To identify which categories differed, we conducted Post Hoc Dunn's tests with holm's adjustment. All pairwise contrasts were statistically significant (small vs.\ medium: \(p < 0.001\); small vs.\ large: \(p < 0.001\); medium vs.\ large: \(p < 0.001\)), indicating that participation ratios differ between every pair of project-size categories. To further quantify the magnitude of pairwise differences, we computed nonparametric effect sizes using Cliff's Delta with 95\% confidence intervals. All comparisons exhibited large effect sizes: small vs.\ medium (\(\delta = 0.9411\), 95\% CI \([0.9164, 0.9612]\)), small vs.\ large (\(\delta = 0.9969\), 95\% CI \([0.9944, 0.9987]\)), and medium vs.\ large (\(\delta = 0.9012\), 95\% CI \([0.8818, 0.9198]\)). These consistently large effect sizes indicate substantial and practically meaningful differences in participation ratios across all project-size categories.

To assess whether the above findings are sensitive to how project size is operationalized, we repeated the analysis using contributor-based definition (Small: 1-20 contributors, Medium: 21-100, Large: 100+). The Kruskal-Wallis H-test again indicated a statistically significant difference in participation ratios across the three bins ($H = 1444.36$, $p < 0.001$), with an even larger effect size ($\eta^2 = 0.7340$). Additionally, Post hoc Dunn's tests with holm's correction showed that all pairwise contrasts were statistically significant (small vs.\ medium: \(p < 0.001\); small vs.\ large: \(p < 0.001\); medium vs.\ large: \(p < 0.001\)), indicating that participation ratios differ between every pair of contributor-size bins. To complement these findings, we also estimated pairwise effect magnitudes using Cliff's Delta with 95\% confidence intervals. The results indicate large effects for all comparisons: small vs.\ medium (\(\delta = 0.8764\), 95\% CI \([0.8521, 0.8985]\)), small vs.\ large (\(\delta = 0.9882\), 95\% CI \([0.9809, 0.9941]\)), and medium vs.\ large (\(\delta = 0.7867\), 95\% CI \([0.7494, 0.8247]\)). Collectively, these effect sizes reinforce the presence of pronounced differences in participation ratios among the contributor-based project size groups

Together, these results provide strong evidence that project size is strongly related to contributor participation in agentic workflows.

\begin{figure}[t]
    \centering
    \includegraphics[width=\linewidth]{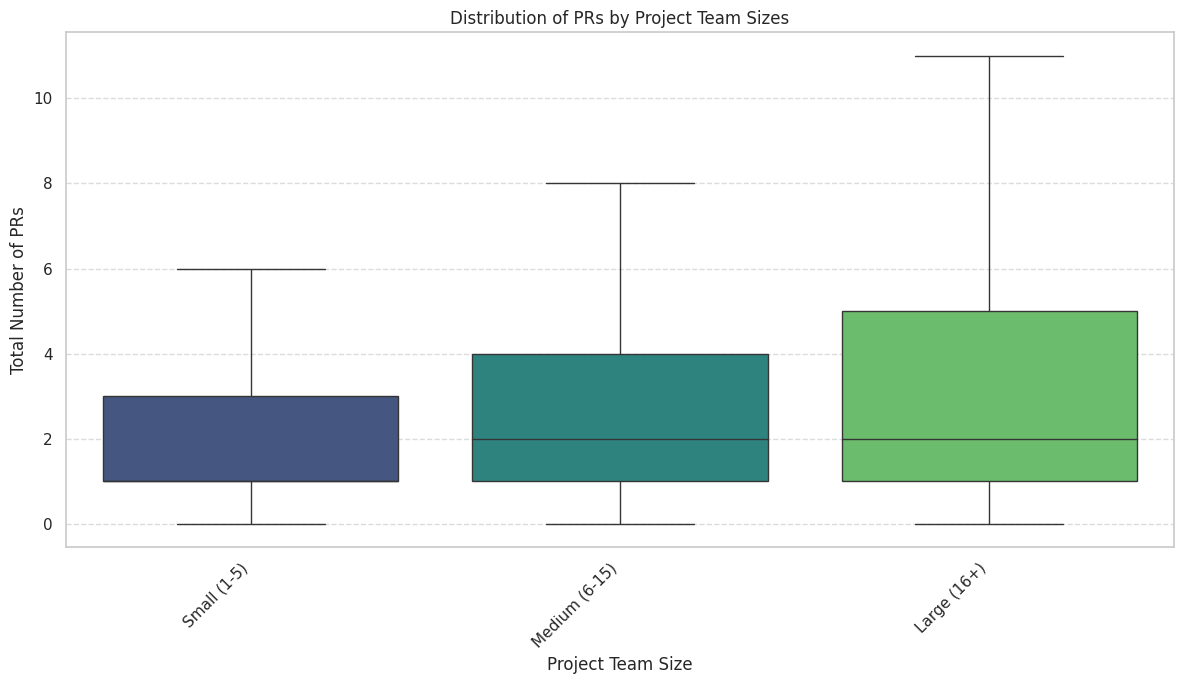}
    \caption{Distribution of agentic PR counts across small, medium, and large project teams (without showing outliers)}
    \label{fig:team_size_boxplot}
    \vspace{-5mm}
\end{figure}

Figure~\ref{fig:team_size_boxplot} shows the distribution of agentic PR counts across small, medium, and large project teams. Across all three project-size categories, the median number of agentic PRs remains low, ranging from one to two PRs per repository during the three-month observation period. This finding indicates that intensive use of agentic coding tools remains uncommon for the typical repository.

The figure also reveals differences in the variability of agentic PR activity across project sizes. Large projects exhibit the widest distribution of PR counts, followed by medium-sized projects, while small projects show the narrowest distribution. Thus, although participation ratios are highest among small projects, the number of agentic PRs generated by individual repositories varies more substantially among larger projects.

At the same time, mean PR counts differ markedly across project sizes. Small projects generate an average of 50.2 agentic PRs per repository, compared with 5.6 and 6.7 for medium-sized and large projects, respectively. The substantial gap between the mean values and the distributions shown in Figure 3 suggests that agentic PR activity is highly skewed, particularly among small projects, where a relatively small number of repositories account for a disproportionate share of agentic PRs.

Taken together, the participation-ratio and PR-count results reveal a consistent pattern. Small projects involve a larger proportion of their contributors in agentic workflows and include some highly active repositories that generate large numbers of agentic PRs. In contrast, medium-sized and large projects engage a smaller fraction of their contributors in agentic workflows, although they exhibit greater variability in PR counts across projects.\\

\begin{summary}
    \textbf{RQ1 Summary:}
    Agentic coding tool adoption varies substantially across project sizes. Small projects involve a larger proportion of their contributors in agentic workflows and achieve the highest average levels of agentic PR activity. However, adoption remains modest for the typical repository across all project-size categories, with median agentic PR counts of only one to two PRs during the three-month observation period. These patterns are robust to alternative definitions of team size based on total number of contributors.
\end{summary}

\subsection{RQ2: Project-Level Agentic PR Productivity}
\begin{figure}[t]
    \centering
    \includegraphics[width=\linewidth]{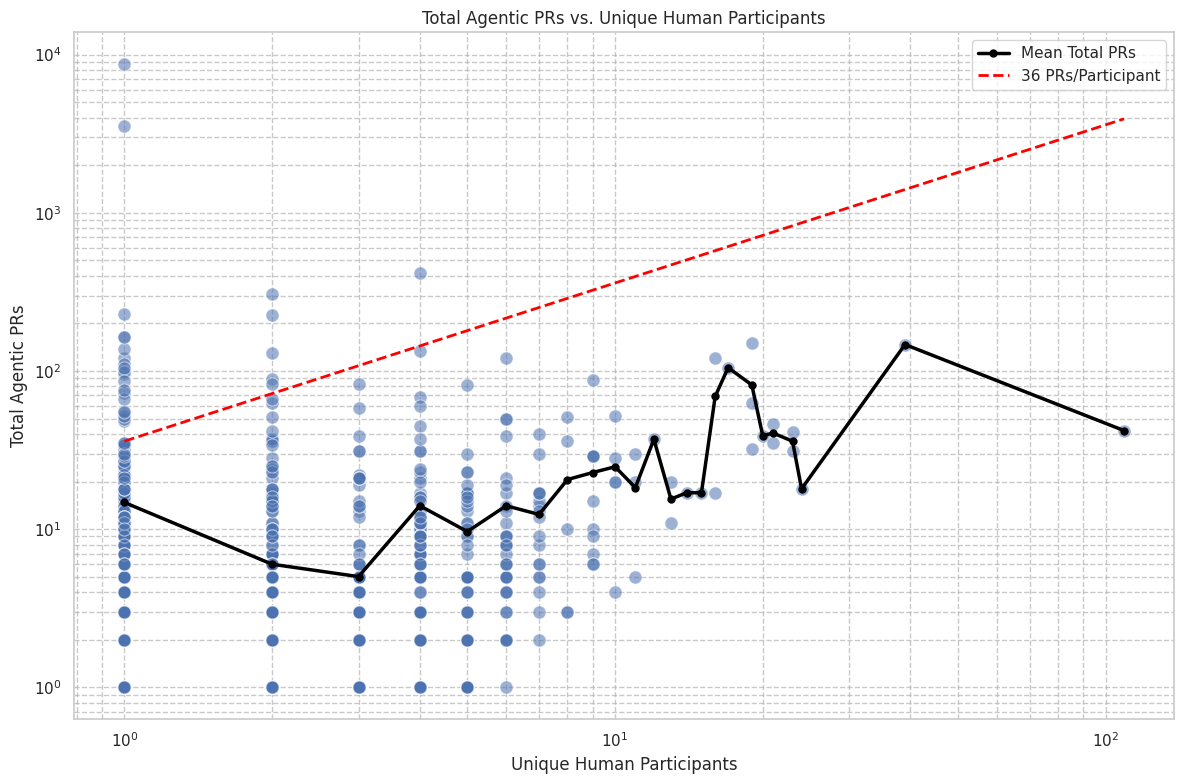}
    \caption{Distribution of total agentic PRs by the number of unique human participants per project. The black line shows the mean PR count, and the red dashed line marks the human PR productivity target, assuming the median human PR productivity is 36 PRs per developer in three months.}
    \label{fig:pr_productivity}
 
\end{figure}

To examine project-level agentic PR productivity, we measure the number of agentic PRs generated per human participant during the three-month observation period. This measure serves as a proxy for the volume of agentic contributions generated and managed by participating humans. Figure~\ref{fig:pr_productivity} plots the number of agentic PRs against the number of human participants for each repository.

As a reference point, Figure~\ref{fig:pr_productivity} includes a reference line at 36 PRs per developer over a three-month period, derived from prior industry observations~\cite{worklytics2025benchmarks}. We use this line only as a contextual reference point, not as a universal productivity standard. Projects above this reference line generated a comparatively high volume of agentic PRs per participating human, whereas projects below the line show lower relative levels of agentic PR activity.

Figure~\ref{fig:pr_productivity} shows substantial variation in project-level agentic PR productivity. Projects with only one or a few human participants exhibit the greatest variation, ranging from a small number of agentic PRs to several hundred. In contrast, projects with larger participant pools tend to cluster more closely around lower productivity values, indicating that high levels of agentic PR activity are relatively uncommon in projects with many participating humans.

Overall, most projects fall below the benchmark line. Only 25 of the 2,361 projects (1\%) exceed the reference threshold of 36 agentic PRs per participant during the three-month observation period. This result suggests that, although agentic coding tools are actively used across many projects, relatively few projects generate agentic PRs at levels comparable to the benchmark productivity value.

At the same time, the existence of projects far above the reference line demonstrates that high levels of agentic PR productivity are achievable in some settings. The wide dispersion observed among smaller projects indicates that the ability to generate and manage large volumes of agentic PRs may depend on project-specific factors such as development practices, contributor engagement, or workflow design. Investigating these factors represents an important direction for future research.\\

\begin{summary}
\textbf{RQ2 Summary:}
Project-level agentic PR productivity varies substantially across projects' participants. While a small number of projects exceed the 36-PR reference line during the three-month observation period, most projects remain below it. High levels of agentic PR productivity are concentrated in a relatively small subset of projects, particularly those with fewer participating humans.
\end{summary}

\subsection{RQ3: Human-Agent Collaboration Patterns Across Project Team Sizes}
\begin{figure}[t]
    \centering
    \includegraphics[width=\linewidth]{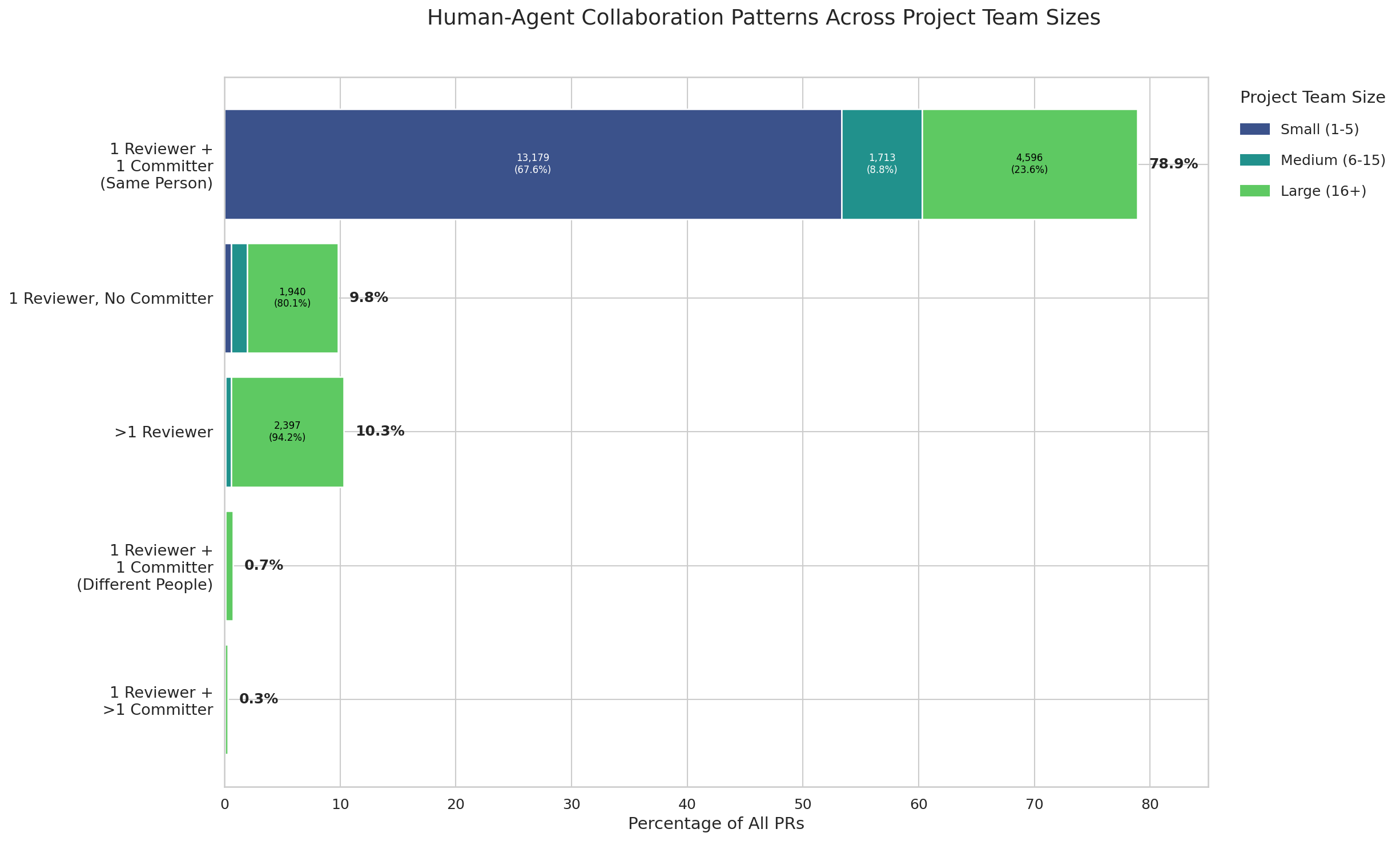}
    \caption{Distribution of human-agent collaboration patterns across project contributor size}
    \label{fig:collaboration-patterns}
   \vspace{-5mm}
\end{figure}

To understand how humans collaborate with coding agents, we classify each agentic PR according to the number and roles of human participants involved in reviewing, modifying, and integrating the agent's contribution. Figure~\ref{fig:collaboration-patterns} summarizes the distribution of collaboration patterns across all agentic PRs.

The most common collaboration pattern is \emph{1 Reviewer + 1 Committer (Same Person)}, accounting for 19,488 PRs (78.9\%). In this workflow, a single developer both reviews and modifies the agent-generated contribution. The second most common pattern is \emph{1 Reviewer, No Committer}, accounting for 2,476 PRs (9.8\%). Together, these two single-human workflows account for 88.7\% of all agentic PRs, indicating that human-agent collaboration is overwhelmingly centered on individual oversight rather than team-based coordination.

In contrast, collaboration patterns involving multiple humans are comparatively rare. PRs involving two or more human participants account for only 11.3\% of all agentic PRs. Among these patterns,  different individuals participate in reviewing and committing the PR. However, each multi-human pattern represents only a small fraction of the overall dataset.

Figure~\ref{fig:collaboration-patterns} further examines collaboration patterns across project-size categories. Small projects are dominated by single-human workflows, particularly the \emph{1 Reviewer + 1 Committer (Same Person)} pattern. Medium-sized and large projects exhibit a more diverse set of collaboration structures and a higher proportion of multi-human workflows. Nevertheless, even in large projects, single-human oversight remains the predominant form of human-agent collaboration.

These findings suggest that current agentic PR workflows place substantial responsibility on individual developers. Rather than replacing human oversight, coding agents appear to operate within development processes that continue to rely heavily on human review and integration activities. Although larger projects show greater evidence of distributed collaboration, team-based oversight remains relatively uncommon in the current generation of agentic software engineering practices.\\

\begin{summary}
\textbf{RQ3 Summary:}

Human-agent collaboration is concentrated in individual developer oversight. In 78.9\% of agentic PRs, one developer both reviews and commits with the agent's contribution, while multi-human collaboration occurs in only a small minority of cases. Larger projects exhibit somewhat more distributed review practices, but single-human workflows remain the dominant collaboration model across all project sizes.
\end{summary}
\section{Discussion}

Our findings provide a project-level perspective on how agentic coding tools are being adopted and integrated into open-source software development workflows. Across all three research questions, a consistent pattern emerges: although agentic coding tools are present across thousands of projects, their use remains concentrated within a relatively small subset of projects and contributors.

First, project size is strongly associated with agentic tool adoption. Small projects exhibit substantially higher participation ratios and higher average levels of agentic PR activity than medium-sized and large projects. At the same time, participation in agentic workflows remains limited in most projects, with fewer than 20\% of contributors involved in agentic PR activities in more than two-thirds of projects. These findings suggest that agentic coding tools have not yet become a routine part of development workflows for most contributors, even in projects where such tools are present.

Although our results demonstrate clear differences across project sizes, the underlying causes of these differences remain an open question. Future research should investigate how organizational factors, development practices, and project governance influence the adoption of agentic coding tools.

Second, our productivity analysis reveals substantial differences in the volume of agentic PR activity generated per human participant. While a small number of projects exceed the prior industry observation of 36 PRs per participant during the three-month observation period, most projects remain below this threshold. This result indicates that high levels of agentic PR productivity are currently concentrated in a relatively small subset of projects. The observed variation suggests that project-level factors may influence the extent to which repositories are able to generate and manage agentic contributions, although identifying those factors is beyond the scope of the present study.

Third, human-agent collaboration is dominated by a single-human oversight model. More than two-thirds of agentic PRs involve a single developer who both reviews and modifies the agent-generated contribution, while multi-human collaboration patterns account for only a small minority of PRs. Although larger projects exhibit somewhat more distributed review practices, individual oversight remains the dominant form of human-agent collaboration across all project-size categories. These findings indicate that human review and integration continue to play a central role in current agentic software development workflows.

The results suggest that the successful integration of agentic coding tools depends not only on advances in agent capabilities but also on the human and organizational processes that govern their use. As agent-generated contributions become more common, projects may need to adapt review practices, accountability structures, and collaboration processes to support sustainable human-agent software development. These findings provide early empirical evidence on how open-source communities organize human oversight around agentic coding tools at the project-level. Future research should investigate the organizational and technical factors associated with successful agent adoption and examine how human oversight practices evolve as agentic coding tools become more widely used.

\section{Threats to Validity}
Several limitations should be considered when interpreting our findings.

Our study relies on the AIDev-pop dataset, which contains agentic PRs from popular GitHub projects with more than 100 stars. Consequently, our findings may not generalize to less popular projects, private projects, commercial software development environments, or projects that use agentic coding tools differently. In addition, we focus on only three coding agents (GitHub Copilot, OpenAI Codex, and Claude Code) and a three-month observation period, limiting the extent to which the results can be generalized to other tools or future development practices.

Second, our identification of human participants is based on information extracted from commits, reviews, comments, and timeline events. Although we applied keyword-based filtering to remove bot and agent accounts, some automated accounts may remain undetected, while some human accounts may be incorrectly excluded. Such inaccuracies could affect the measured levels of human participation and collaboration.

Third, we use the number of repository contributors obtained through the GitHub API as a proxy for project team size. Contributor counts may not fully reflect the active developers involved in a project's day-to-day maintenance or decision-making processes. As a result, our team-size classifications should be interpreted as an approximate measure of project size rather than a direct measure of organizational structure.

Finally, our productivity analysis operationalizes project-level agentic PR productivity as the number of agentic PRs per human participant during the three-month observation period. Although this measure provides a useful proxy for the volume of agentic contributions generated and managed by participating humans, it does not account for differences in PR size, complexity, quality, or review effort. Therefore, the productivity findings should be interpreted as reflecting relative levels of agentic PR activity rather than comprehensive measures of software development productivity.

Overall, the findings from this study should be interpreted as an empirical snapshot of current human-agent collaboration practices in popular open-source projects during an early stage of agentic coding tool adoption.
\section{Conclusion}
This study analyzed 25,264 agentic pull requests (PRs) from 2,361 open-source GitHub repositories to examine project-level adoption of agentic coding tools, project-level agentic PR productivity, and human-agent collaboration patterns.

Our findings show that adoption is highly uneven across projects. The median number of agentic PRs remains low across all project-size categories, ranging from one to two PRs per repository during the three-month observation period. This finding suggests that, although agentic coding tools have been adopted across thousands of projects, intensive adoption remains concentrated in a relatively small subset of projects. At the same time, small projects (1-5 contributors) exhibit substantially higher participation ratios and average levels of agentic PR activity than medium-sized and large projects, while also involving a larger proportion of their contributors in agentic workflows. In contrast, larger projects adopt agentic tools more selectively and engage only a small fraction of their contributor base in agentic PR activities.

We also find substantial variation in project-level agentic PR productivity. While a small number of projects exceed an industry-reported baseline of 36 PRs per participant during the three-month observation period, most projects remain below this threshold. High levels of agentic PR productivity are therefore concentrated in a relatively small subset of projects.

Furthermore, human-agent collaboration is dominated by a single-human oversight model, in which one developer both reviews and modifies the agent's contributions. Although larger projects show greater evidence of distributed review practices, such workflows remain relatively uncommon.

Because this study captures an early snapshot of agent adoption, future work
should continue to track how adoption patterns evolve over time. Taken together, the findings from this study suggest that the successful integration of agentic coding tools depends not only on advances in agent capabilities but also on the human and organizational processes that govern their use. As agent-generated contributions become more common, projects may need to adapt review processes, accountability structures, and collaboration practices to support sustainable human-agent software development. Future work should continue to investigate the organizational and technical factors associated with successful agent adoption and how human oversight practices evolve as agentic coding tools become more widely adopted in open-source ecosystems.

\balance
\bibliographystyle{ACM-Reference-Format}
\bibliography{references.bib}
\end{document}